\documentclass[pra,twocolumn,superscriptaddress,aps,nofootinbib]{revtex4-1}
\usepackage{amsmath}
\usepackage{amsfonts}
\usepackage{amssymb}
\usepackage{graphicx}
\usepackage[usenames]{color}
\usepackage{soul} %\hl
\usepackage{braket}

\usepackage{color}
\usepackage{hyperref} %typesetting URLs
\newcommand{\ba}{\hat{b}^{}}
\newcommand{\bd}{\hat{b}^\dagger}
\newcommand{\residual}{r}
\newcommand{\decay}{\kappa}
\newcommand{\be}{\begin{equation}}
\newcommand{\ee}{\end{equation}}
\renewcommand{\Re}{\operatorname{Re}}

\begin{document}
\title{Coherence protection in coupled quantum systems}
\author{H.~M.~Cammack}
\affiliation{SUPA, School of Physics and Astronomy, University of
  St.~Andrews, KY16 9SS, U.K.}
  
\author{P.~Kirton}
\affiliation{SUPA, School of Physics and Astronomy, University of
  St.~Andrews, KY16 9SS, U.K.}

\author{T. M. Stace}
\affiliation{Centre for Engineered Quantum Systems, School of Mathematics and Physics, The University of Queensland,
St Lucia, Queensland 4072, Australia}
\author{P.~R.~Eastham}

\affiliation{School of Physics and CRANN, Trinity College Dublin, Dublin 2, Ireland.}

\author{J.~Keeling}
\affiliation{SUPA, School of Physics and Astronomy, University of
  St.~Andrews, KY16 9SS, U.K.}
  
\author{B.~W.~Lovett}
\affiliation{SUPA, School of Physics and Astronomy, University of
 St.~Andrews, KY16 9SS, U.K.}

\begin{abstract}
% Needs to be 600 characters; currently 620
The interaction of a quantum system with its environment causes decoherence, setting a fundamental limit on its suitability for quantum information processing. However, we show that if the system consists of coupled parts with different internal energy scales then the interaction of one part with a thermal bath need not lead to loss of coherence from the other. Remarkably, we find that the protected part can remain coherent for longer when the coupling to the bath becomes stronger or the temperature is raised. Our theory will enable the design of decoherence-resistant hybrid quantum computers.
\end{abstract}
\maketitle

\section{Introduction}
Quantum decoherence describes the process by which a small system becomes entangled with a large environment, such that its phase can no longer be defined locally~\cite{ekert97,breuerpet, schlosshauer05}. Indeed, a single quantum system coupled weakly to a bosonic bath must lose coherence in its energy eigenbasis~\cite{paz99, zurek03}. In some sense the bath makes a `measurement' of the system, and its state becomes mixed.  However, the case of a hybrid quantum system --- i.e. one that is composed of multiple interacting parts --- is much more subtle. When the constituent parts have distinct energy scales, a bath can couple to each part with a different strength, and the decoherence of one part can influence the coherence of other parts.

\begin{figure}[h!]
    \begin{center}
        \includegraphics[angle=0, width=0.45\textwidth]{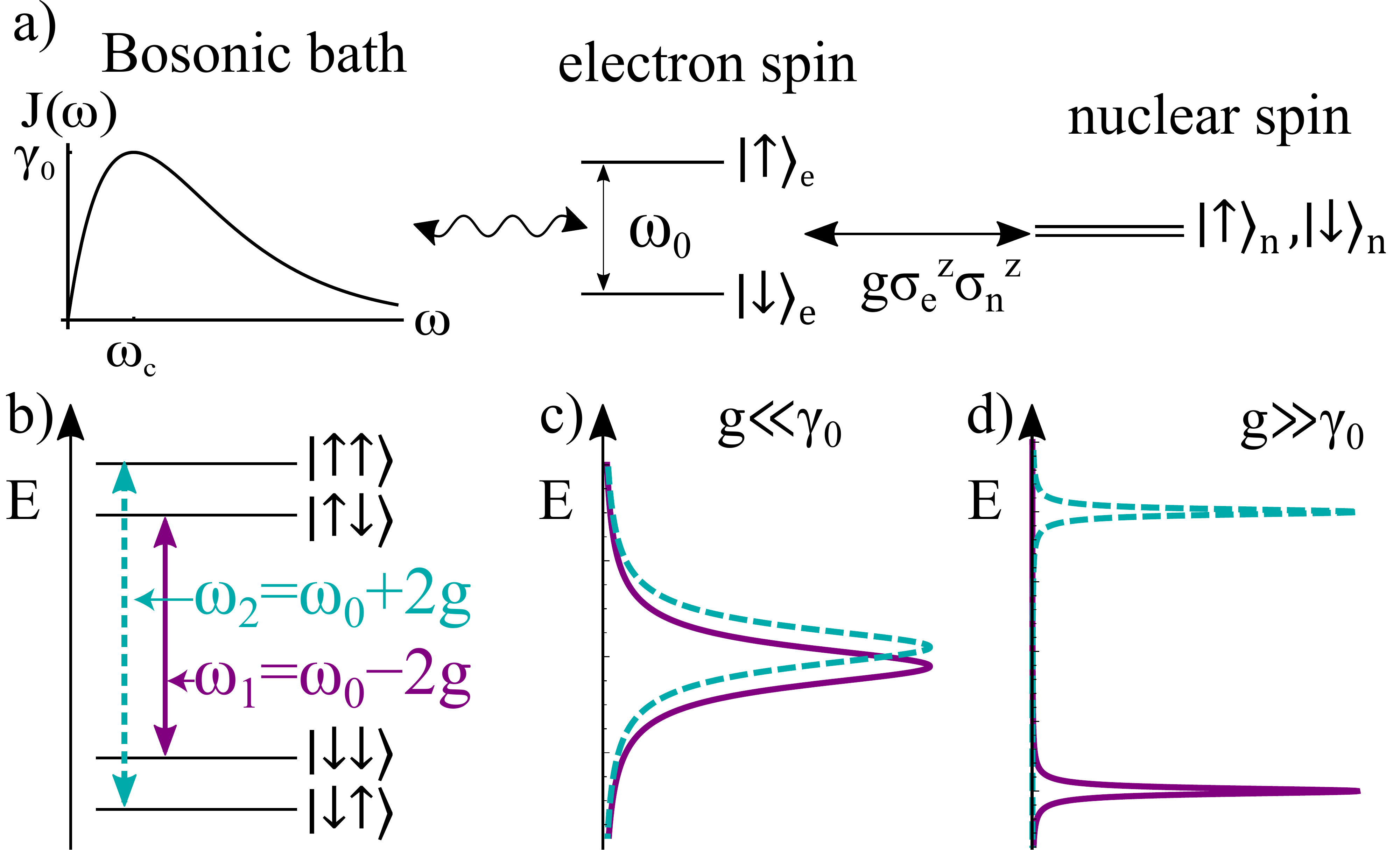}
      \caption{a) A coupled nuclear-electron spin system with electron Zeeman splitting of $\omega_0$ and hyperfine coupling $g$. We neglect the Zeeman splitting of the nuclear spin, and only the electron interacts with a thermal reservoir. b) The eigenenergies of the two-spin system and allowed transitions $\omega_{1,2}$. When the hyperfine coupling is small (c), the transitions overlap and nuclear spin coherences can survive. Increasing the hyperfine coupling (d) resolves the transition lines and coherences are lost.} %more detail?
        \label{spin1spin2}
   \end{center}
\end{figure}

In this paper, we will explore how this interplay of intra-system and system-bath coupling can affect the coherence of one part of a coupled system. In particular, we will explore whether conditions exist under which such coherence can be protected from bath-induced noise.
The system we will study consists of two qubits with energy scales differing by at least one order of magnitude (see Fig.~\ref{spin1spin2}). The theory we discuss is generally applicable, and could apply for example to exciton-electron spin hybrids in quantum dots~\cite{atature07, ramsay08}, molecular systems~\cite{schaffry10} or defects in crystals~\cite{neumann10}, or to superconducting qubits coupled to spins~\cite{nori13} or defect centres~\cite{marcos10}.  However, we will base our argument on the concrete example of a coupled spin-half electron and spin-half nucleus~\cite{Morton2011}. Hybrid systems have been proposed as quantum computing devices~\cite{Kurizki2015}, for example represented by donors in silicon~\cite{Morley2012,Zhao2013}, that are able to perform more efficiently than if only one qubit type is used. Electron spins have rapid manipulation times, and these can be coupled to optical photons or superconducting qubits for measurement or entanglement~\cite{Morton2011,Kurizki2015}, and coupling between electron and nuclear spins provides access to quantum memory with ultra long nuclear coherence times~\cite{Zhong2015}. 

As an illustrative example, we will use a simplified description in which the electron spin decays by emitting energy into a large environment (see for example~\cite{bertet16} for a situation in which photon emission dominates spin decay), but the nucleus only interacts with the electron. This coupling takes a form such that the transition energy of the electron has a different value for each of the two nuclear spin states. We might then expect that the rate of electron spin decay would have a profound effect on whether or not the environment destroys coherent superpositions of nuclear states by projecting the nucleus onto one of its energy eigenstates. For a sufficiently slow decay at low temperature, the different nuclear spin states would have resolved lines in the electron spin emission spectrum, and we will argue this leads to a loss of nuclear coherence (see Fig.~\ref{spin1spin2}). Conversely, we will investigate whether a fast electron spin decay is able to preserve nuclear spin coherence. We will finally explore the impact of increasing temperature on these findings.

We first discuss the behavior at zero temperature (Sec.~\ref{sec:zeroT}).  In this limit, the problem can be solved exactly.  To extend our discussion to finite temperatures, we will in Sec.~\ref{sec:MQME} introduce an approximate Markovian quantum master equation (MQME). By comparing its predictions for zero temperature with the exact solution we will check the appropriateness of this equation. Using the MQME, the dependence of the long-time nuclear coherence on the decay rate of the electron spin will be presented and discussed, and the high temperature behaviour will be checked by comparing with a semiclassical model in Sec.~\ref{sec:semimod}. In Sec.~\ref{discussion} we will suggest an experimental setup that would be capable of testing our predictions.

\section{Model and Exact Solution at Zero Temperature}
\label{sec:zeroT}

Our model system is depicted in Fig.~\ref{spin1spin2} and is described by a total 
Hamiltonian $H=H_S + H_{SB} + H_{B}$:
\begin{align}
H_S &=  \frac{\omega_0 }{2}\hat\sigma_e^z + g\hat\sigma_e^z \hat\sigma_n^z, \nonumber \\
H_{SB}&=\sum_k c_{k} (\hat\sigma_e^+\ba_{k} +\hat\sigma_e^- \bd_{k}), \nonumber \\
H_{B}&=\sum_k \nu^{}_k \bd_{k} \ba_{k}, 
\end{align}
where $\hat\sigma_{e(n)}^z$ is the Pauli $z$ operator for the electron (nucleus), and $\hat\sigma_e^{+(-)}$ is the electron spin raising (lowering) operator.
$\omega_0$ is the electron Zeeman splitting and $g$ is the hyperfine coupling strength. We ignore the nuclear Zeeman splitting which is negligible on this scale, and including it would have no impact on our conclusions. Similarly we consider only the Zeeman ($z$) part of the electron-nuclear coupling.  The effect on the dynamics of any spin-flip  ($x,y$) part of the electron-nuclear coupling is strongly suppressed due to the large difference between electron and nuclear Zeeman energies.  The environment, which couples only to the electron, is modelled as a collection of quantum harmonic oscillators where a quantum in mode $k$ is created with $b_k^\dagger$.  The effects of the environment are characterized by the spectral density $J(\nu) = \pi \sum_k |c_k|^2 \delta(\nu-\nu^{}_k)$. 
When calculating numerical
results, we use a spectral density of Ohmic form,
\be
\label{eq:spec}
J(\nu)=\gamma_0\left(\frac{\nu}{\xi_c}\right)^s\exp\left(s-\frac{s\nu}{\xi_c}\right)
,\ee
where $\xi_c$ is a cutoff frequency and $\gamma_0$ is the peak spectral density. $s$ describes the ohmicity of the bath and unless otherwise stated we will assume an Ohmic bath, $s=1$. We will also choose $\xi_c=\omega_0$ unless explicitly stated, since the bath is approximately flat around the peak at $\nu=\xi_c$ --- though our conclusions are valid for any choice. In the following calculations we will switch to the interaction picture with respect to $H_0=H_S + H_B$.

We first present the exact (non-Markovian) zero-temperature solution using the Wigner-Weisskopf method \cite{Wigner1930,Vacchini2010}. Since the bath is at $T=0$ and the Hamiltonian is number conserving we may work in the subspace where there is a single excitation in either the electron spin or bath. Therefore, a general state at time $t$ can be written:
\begin{align}
|\Psi(t)\rangle=&+a_1(t)\mid\uparrow\downarrow\rangle\otimes|0\rangle_B+a_2(t)\mid\uparrow\uparrow\rangle\otimes|0\rangle_B \nonumber \\
&+\sum_k \alpha_{1,k}(t)\mid\downarrow\downarrow\rangle\otimes|k\rangle_B+\alpha_{2,k}(t)\mid\downarrow\uparrow\rangle\otimes|k\rangle_B.
\end{align}
Here the system state is given in terms of the eigenstates of $\hat\sigma_z$ ($\mid\uparrow\rangle, \mid\downarrow\rangle$) and with the electron spin specified first and the nuclear spin second. The state $\ket{0}_B$ is the vacuum state of the bath and $\ket{k}_B$ refers to the $k$th bath mode containing the excitation. We then use the Schr\"{o}dinger equation to obtain differential equations for state coefficients, $\alpha_{1,k}, \alpha_{2,k}, a_1$ and $a_2$. After Laplace transforming we find the equation for the states with an electronic excitation:
\begin{equation}
\tilde{a}_m(s) \big( s + f(s-i\omega_m)\big) = a_m(0),
\label{eqLaplaceT}
\end{equation}
where we denote Laplace transformed functions with a tilde. $m \in \{1,2\}$ and the transition frequencies are $\omega_1=\omega_0-2g$, $\omega_2=\omega_0+2g$ (as depicted in Fig.~\ref{spin1spin2}), and 
\begin{equation}
f(s)\equiv \sum_k \frac{ |c_k|^2}{s+i\nu^{}_k} = \frac{1}{\pi} \int_0^\infty d\nu \frac{J(\nu)}{s+i\nu}.
\label{eqfs}
\end{equation}
Eq.~\eqref{eqLaplaceT} further allows us to write:
\begin{equation}
\tilde{\alpha}_{m,k}(s)=-\frac{ic_k}{s}\frac{a_m(0)}{s+i(\omega_m-\nu^{}_k)+f(s-i\nu^{}_k)}.
\label{eqalphaLap}
\end{equation}

We can now find the coherence between nuclear spin levels when the electron spin has decayed to its ground state. The element of the density matrix denoting such a coherence with the electron in its ground state is written $\rho_{\downarrow\downarrow,\downarrow\uparrow}(t)=\langle\downarrow\downarrow\mid \rho (t)\mid \downarrow\uparrow\rangle= \sum_k \alpha_{1,k} (t) \alpha_{2,k} ^*(t)$. Using the inverse Laplace transform of Eq.~\eqref{eqalphaLap} yields:
\begin{align}
\rho_{\downarrow\downarrow,\downarrow\uparrow}(t)=&-\frac{1}{4\pi^3} \int_{\mathcal B} ds_1 \int_{\mathcal B} ds_2 e^{(s_1+s_2)t} a_1(0)a^*_2(0)\nonumber \\
\times\int_0^\infty d\nu &J(\nu) \frac{1}{s_1 \big( s_1+i(\omega_1-\nu)+f(s_1-i\nu)\big)} \nonumber \\
&\times \frac{1}{s_2 \big( s_2-i(\omega_2-\nu)+f^*(s_2-i\nu)\big)},
\label{eqinvLaprho34}
\end{align}
where $\mathcal{B}$ denotes the Bromwich contour.
Each of the $s$-integrals contains a pole at zero, and other structure (poles or branch cuts) elsewhere in the complex plane ~\cite{PhysRevLett.111.180602,Eastham16}.  In the long time limit, the only surviving contributions are from the $s=0$ poles. To evaluate the residue at $s=0$, we must find $f(-i\nu+0_+)$. Using Eq.~\eqref{eqfs}:
\begin{equation}
f(-i\nu+0_+)=J(\nu)-\frac{i}{\pi}\mathcal{P}\int_0^\infty dx \frac{J(x)}{x-\nu} \equiv\Gamma(\nu),
\label{eqgamma}
\end{equation}
leading to:
\begin{align}
\label{rhoexactSS}
\lim_{t\to\infty} \rho_{\downarrow\downarrow,\downarrow\uparrow}(t)&=a_1(0)a^*_2(0)\frac{1}{\pi} \\
&\times\int_0^\infty d\nu \frac{J(\nu)}{(\omega_1-\nu-i\Gamma(\nu))(\omega_2-\nu+i\Gamma^*(\nu))}.\nonumber
\end{align}

\begin{figure}[h!]
\begin{center}
\includegraphics[width=0.45\textwidth]{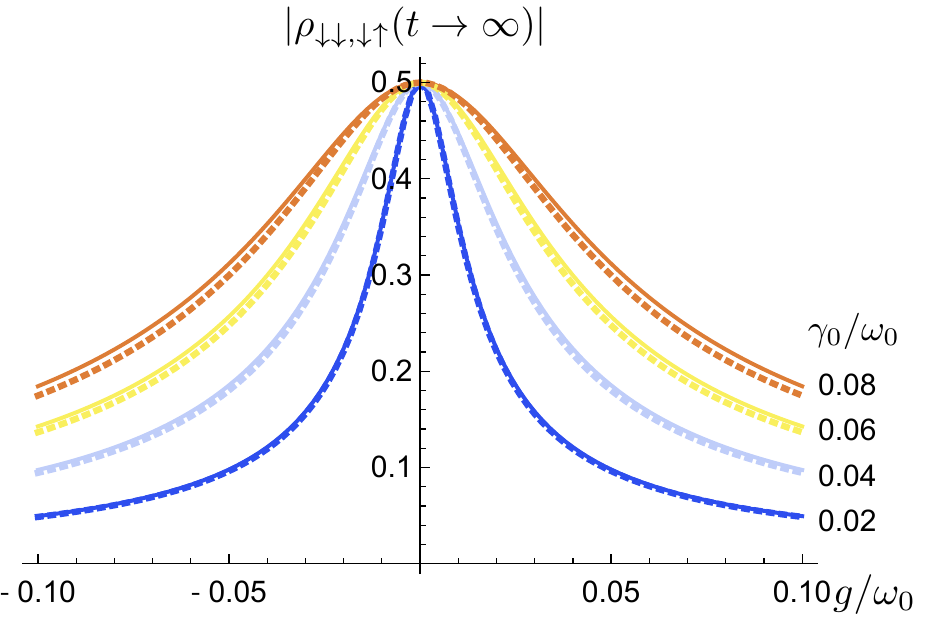}
\caption{The magnitude of the steady state coherence in the lower electron state, $\rho_{\downarrow\downarrow,\downarrow\uparrow}$, at zero temperature as a function of the spin-spin coupling, $g$. The bath has an Ohmic spectral density peaked on resonance with the bare electron splitting $\omega_0$ and peak rate $\gamma_0$. Solid lines represent the exact solutions, whilst dashed lines show the solutions using the Born-Markov approximation.}
\label{figexactBM}
\end{center}
\end{figure}

Following preparation in the initial state $|\psi(0)\rangle = \ket{\uparrow}(\ket{\downarrow}+\ket{\uparrow})/\sqrt{2}$, we plot the exact steady-state coherence in  Fig.~\ref{figexactBM}. %for an Ohmic spectral density with peak at the bare electron splitting ($\omega_0=\xi_c$).
When $g$ is small compared with the typical electronic decay rate $\gamma_0$, nuclear coherence is preserved close to its maximum value of 1/2 throughout the electronic decay process. Increasing $g$ causes a reduction and eventual loss of nuclear coherence, with the effect more pronounced for smaller electron-bath coupling rates.  
Increasing coupling to the bath broadens the range of $g$ over which coherence is preserved, so that at fixed $g$, increasing damping causes increased coherence.

To see why this is, consider how the initial state $|\psi(0)\rangle$ reaches equilibrium.  Our theory applies to any bosonic bath but to give a concrete illustration, we imagine that the bath corresponds to photon modes, and so describes photon emission as the electron relaxes to its ground state. The emitted photon spectrum would have two peaks, one centred on each transition frequency $\omega_0\pm 2g$, and corresponding to the two possible nuclear spin states when the decay occurred. Each peak has a width proportional to the transition rate. If the separation between the two transitions is greater than the transition rates, there is little overlap between the $\mid\uparrow\rangle_{n}$ and $\mid\downarrow\rangle_{n}$ frequency distributions, as shown in Fig.~\ref{spin1spin2}. Measuring the energy of the photon would allow us to know which transition occurred, and thus which state the nuclear spin is in -- any coherence is lost, even if the measurement outcome is not recorded. If the frequency distributions of photons emitted from the two different transitions overlap sufficiently, then measuring the frequency of the bath excitation does not allow one to determine which decay occurred, and so the nuclear spin remains in a superposition.

Having established and understood the behaviour at zero temperature, in the next section we extend our study to finite temperature using both quantum and semiclassical models.

\section{Approximate treatments at finite temperature}

\subsection{Markovian Quantum Master equation}
\label{sec:MQME}

In order to explore non-zero temperature behaviour, we now derive a MQME, an approximate but fully quantum mechanical model, for the same system. We will check the validity of the MQME by comparing to the exact approach just described at zero temperature. The resulting equation will then allow us to extend our predictions to finite temperature. When deriving a MQME, several key approximations are typically made. Two of these are the Born approximation and the Markov approximation, both of which are correct to second order in the system-environment coupling~\cite{breuerpet}.   In calculations like these, an additional secular approximation is often made, where dissipator terms that are off-diagonal in the energy basis are neglected. This assumption is typically justified in quantum-optical systems, where it is also known as the rotating wave approximation \cite{breuerpet}, and leads to a Lindblad form for the Liouvillian~\cite{Lindblad1976,Dumcke1979,Jeske2015,Pearle2012}. However,  for systems with closely spaced transition frequencies such as the coupled nuclear-electron spins, secularization, i.e.\ discounting off-diagonal terms, is not justified. Indeed, secularization predicts that the nuclear coherences will always be zero following electron spin decay, except at $g=0$, where the nuclear spin is completely isolated anyway. We know from the exact solution that nuclear coherence can be preserved, and we must therefore avoid secularization here.  As discussed elsewhere~\cite{Jeske2014a,Eastham16}, a microscopically derived Bloch-Redfield equation can yield accurate and physical results for the steady state in such cases.

The Bloch-Redfield equation for the system density matrix $\rho(t)$ following the Born and Markov approximations is~\cite{breuerpet}:
\begin{multline}
\frac{d \rho}{dt}  =  \!\!\!\! \sum_{i,j=1,2} \!\!\!\!\!
e^{i(\omega_i-\omega_j)t} \Big( \Gamma^\uparrow(\omega_i)D^{\uparrow}_{ij}[\rho]+ \Gamma^{\uparrow\ast}(\omega_j)D^{\uparrow\dagger}_{ji}[\rho] \\
+  \Gamma^\downarrow(\omega_j)D^{\downarrow}_{ji}[\rho]+ \Gamma^{\downarrow\ast}(\omega_i) D^{\downarrow\dagger}_{ij}[\rho]\Big),
\label{eqMarkQME}
\end{multline}
where we have defined $D^{\uparrow}_{ij}[\rho]\equiv A^\dagger_i\rho A_j - A_jA^\dagger_i\rho$, $D^{\downarrow}_{ij}[\rho]\equiv A_i\rho A^\dagger_j - A^\dagger_jA_i\rho$ and the transition operators are \\$A_1=\mid\downarrow\downarrow\rangle\langle\uparrow\downarrow\mid$ and $A_2=\mid\downarrow\uparrow\rangle\langle\uparrow\uparrow\mid$. Here we have introduced the generalizations of the quantity $\Gamma(\omega)$ (see Eq.~\ref{eqgamma}) to non-zero temperatures:
\begin{equation}
  \begin{aligned}
    \Gamma^\downarrow(\omega)&=
    J(\omega)[n(\omega)+1] -\frac{i}{\pi}  \mathcal{P}\!\int_{0}^\infty\frac{J(\nu)[n(\nu)+1]}{\nu-\omega} d\nu\ ,
    \\
    \Gamma^{\uparrow}(\omega)&=
    J(\omega)n(\omega) 
    + \frac{i}{\pi} \mathcal{P}\!\int_{0}^\infty\frac{J(\nu)n(\nu)}{\nu-\omega} d\nu\ ,
  \end{aligned}
  \label{eqgamma2}
\end{equation}
where $n(\omega)=[e^{\hbar \omega/k_B T}-1]^{-1}$ is the Bose-Einstein distribution function, so that at zero temperature $\Gamma^\uparrow(\omega)=0$.

Solving Eq.~\eqref{eqMarkQME} yields the following Born-Markov approximated expressions for the nuclear coherences (written in the interaction picture) when the electron is spin up and spin down respectively:
\begin{align}
\label{eqrhoddud}
\rho_{\downarrow\downarrow,\downarrow\uparrow}(t)=&\left[
\residual_{\downarrow\downarrow,\downarrow\uparrow}( e^{-\decay_- t}- e^{-\decay_+ t})+\rho_{\downarrow\downarrow,\downarrow\uparrow}(0)e^{-\decay_+ t} \right], \\
\rho_{\uparrow\downarrow,\uparrow\uparrow}(t)=&
\left[ \residual_{\uparrow\downarrow,\uparrow\uparrow}  (e^{-\decay_- t}-e^{-\decay_+ t})+\rho_{\uparrow\downarrow,\uparrow\uparrow}(0)e^{-\decay_+ t}\right]e^{-4igt},\nonumber
\end{align}
where, defining $\gamma^{\uparrow(\downarrow)} =\Gamma^{\uparrow(\downarrow)}(\omega_1)+\Gamma^{{\uparrow(\downarrow)}\ast}(\omega_2)$,
we have:
\begin{align}
&\decay_{\pm} =\frac{1}{2}\left[ \gamma^\uparrow +\gamma^\downarrow -4ig \pm \sqrt{(\gamma^\uparrow+\gamma^\downarrow-4ig)^2+16ig\gamma^\uparrow}
\right],  \nonumber \\
&\residual_{\downarrow\downarrow,\downarrow\uparrow} =\frac{\gamma^\downarrow\rho_{\uparrow\downarrow,\uparrow\uparrow}(0)-(\gamma^\uparrow -\decay_+)\rho_{\downarrow\downarrow,\downarrow\uparrow}(0)}{\decay_+ -\decay_-}, \nonumber \\
& \residual_{\uparrow\downarrow,\uparrow\uparrow} = \residual_{\downarrow\downarrow,\downarrow\uparrow}\Big(\frac{\gamma^\uparrow-\decay_-}{\gamma^\downarrow} \Big).
\label{eqkappa}
\end{align}
In the dynamics predicted by Eq.~\eqref{eqrhoddud} there exist two characteristic decay timescales: a fast one governed by $1/|\Re\{\decay_{+}\}|$ and a slow one corresponding to $1/|\Re\{\decay_{-}\}|$. At zero
temperature, $\gamma_\uparrow=0$, we find
$\decay_\pm = [\gamma^\downarrow - 4 ig](1\pm1)/2$, so the slow timescale extends to infinity. \footnote{We note that $\kappa_-(T=0)\equiv 0$ is strictly only true for the $z$ coupling model we consider, neglecting the suppressed $xy$ coupling between electron and nucleus.  If that term is included, a very small but non-zero decay survives even at zero temperature. Therefore the nuclear spin coherence does eventually decay in the isotropic coupling case on a very slow timescale, but a quasi steady state exists before this happens, whose properties are essentially identical to the $z$ coupling steady state.}  Hence, following initialization in state $\ket{\psi(0)}$ we expect a steady state coherence $|\rho_{\downarrow\downarrow,\downarrow\uparrow}(t\to \infty)|=\residual_{\downarrow\downarrow,\downarrow\uparrow}$. This quantity is plotted as dashed lines together with the prediction of the exact solution in Fig.~\ref{figexactBM}. The agreement between the two theories is excellent, so long as the electron-bath coupling strength is not too large ($\gamma_0 \ll \omega_0$). 
From the MQME at we are also able allows us to extract the half width at half maximum $\Delta g_{\rm HWHM}$ of the curves in Fig.~\ref{figexactBM}: if we neglect principal value contributions and assume 
$J(\omega_1)=J(\omega_2)=\gamma_0$ then  $\Delta g_{\rm HWHM}/\omega_0 = \gamma_0\sqrt{3}/(2\omega_0)$.

Having established the validity of the MQME we are able to use it to extend our analysis to finite temperature. Here, $\Re\{\decay_{-}\}$ is not zero, but nonetheless the fast and slow timescales may be significantly different --- and then there can exist a quasi steady state (QSS) once the faster decay process has occurred. In the QSS we find $\rho_{\downarrow\downarrow,\downarrow\uparrow}^{\text{qss}}=\residual_{\downarrow\downarrow,\downarrow\uparrow}$ and $\rho_{\uparrow\downarrow,\uparrow\uparrow}^{\text{qss}}=\residual_{\uparrow\downarrow,\uparrow\uparrow}$. 

\begin{figure}[h!]
\begin{center}
\includegraphics[width=0.5\textwidth]{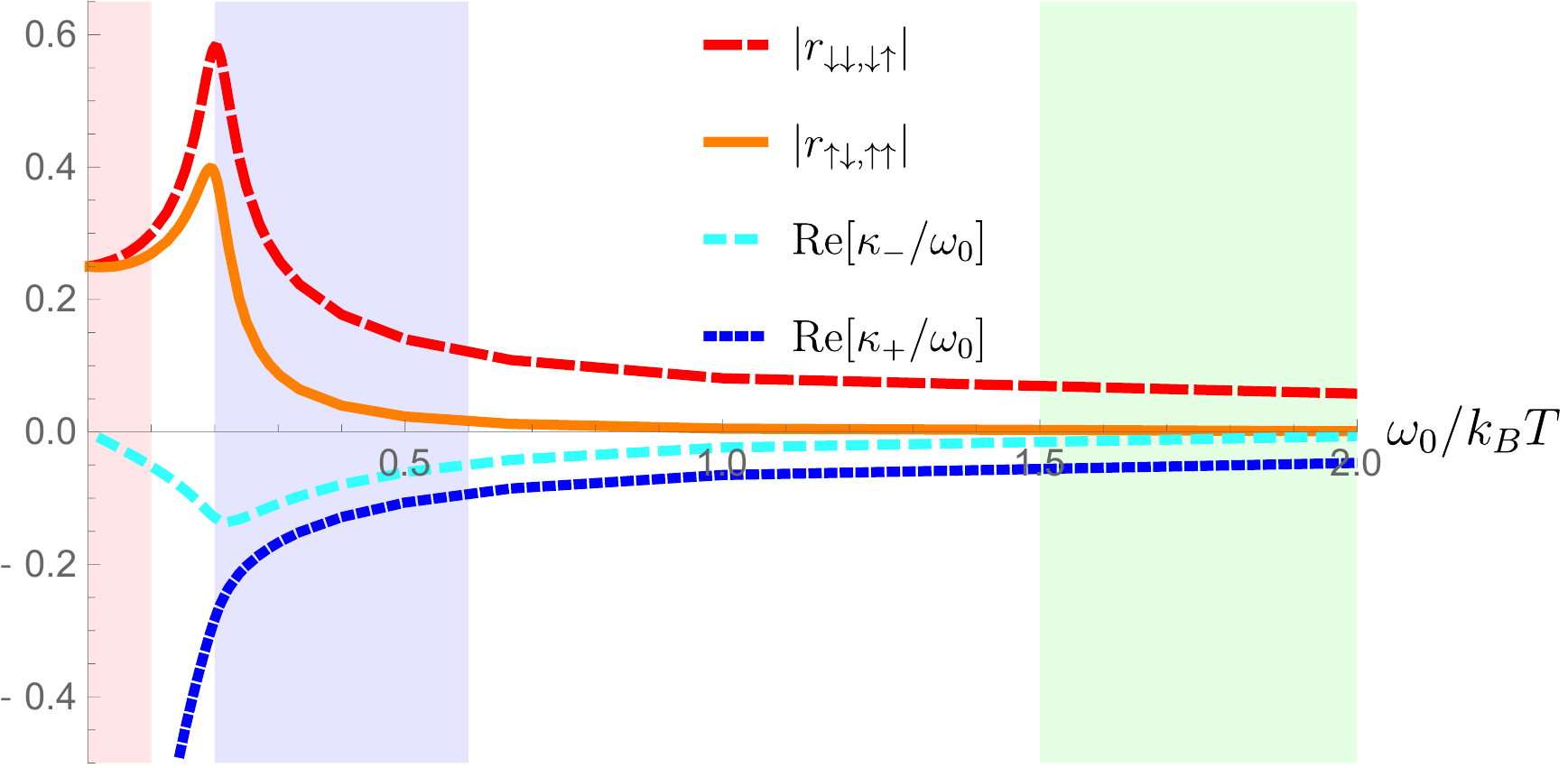}
\caption{The quasi steady state nuclear coherences for the upper (orange) and lower (red, long dashed) electron state, plotted as a function of inverse temperature for $g=0.1\omega_0$, $\gamma_0=0.02\omega_0$. Also plotted are the two characteristic decay rates, $\Re\{\decay_\pm\}$. A quasi steady state is only observed when $\Re\{\decay_-\}$ is sufficiently small. The three coloured areas of background indicate three of the regimes of behaviour discussed in the text: Green (far right shaded area) is low but finite temperature, in which there is a QSS; Blue (middle shaded area) is at an intermediate temperature where there is no QSS; Red (far left shaded area) indicates very high temperature in which a QSS returns.}
\label{kappa2}
\end{center}
\end{figure}

The magnitude of the QSS coherences are plotted in Fig.~\ref{kappa2} as a function of temperature, and there are four regimes of behavior. First, at zero temperature we know $\Re\{\decay_-\}=0$, but we have chosen parameters such that the emission spectrum has partially resolved peaks and so only partial nuclear coherence survives in the steady state.
At low but finite temperature (the green shaded region in the plot), a QSS exists: the electron initially decays much as before, revealing information about the nuclear spin state. However, now there are slow photon absorption processes, which lead to re-excitation and decay of the electron. Each time another decay occurs, more nuclear coherence is lost, and eventually it decays to zero on the long timescale $1/\Re\{\decay_-\}$. As the temperature rises still further, a third regime is entered as $\gamma^{\uparrow(\downarrow)}$ becomes comparable to $g$ (this is the blue shaded region in the plot).  Now the electron decays and reexcites on a similar timescale, yet decays are still slow enough that significant information is revealed about the nuclear spin on each decay: now $\Re\{\decay_-\}\sim\Re\{\decay_+\}$. A QSS no longer really exists, and the coherences of the system do not therefore become $\residual_{\downarrow\downarrow,\downarrow\uparrow}$ and $\residual_{\uparrow\downarrow,\uparrow\uparrow}$. 

The most remarkable feature occurs in the fourth regime (the red shaded region), as the temperature is raised even further. Now the electron spin decays quickly primarily through stimulated emission, and the process becomes so fast that in each decay process very little information is revealed about the state of the nuclear spin. Since now $\gamma^{\uparrow(\downarrow)} \gg g$, we may expand $\kappa_\pm$ in powers of $g/(\gamma^\uparrow+\gamma^\downarrow)$.  For a Markovian bath, we find that $\Re[\decay_+] \simeq \gamma^\uparrow+\gamma^\downarrow, \Re[\decay_-]\simeq 4 g^2 /(\gamma^\uparrow+\gamma^\downarrow)$, so the ratio of fast and slow timescales again becomes small and a QSS returns, with coherences at the highest temperatures reaching $|\rho_{\downarrow\downarrow,\downarrow\uparrow}^{\text{qss}}|=|\rho_{\uparrow\downarrow,\uparrow\uparrow}^{\text{qss}}|=1/4$. These two add in phase with one another, and so after tracing out the electron spin, the nuclear spin coherence takes its initial value of 1/2 in the QSS, which is {\it several times larger than its value in the  zero temperature steady state}. This QSS coherence then slowly decays at the rate $\kappa_-$. Thus, we must conclude that {\it heating a hybrid system can delay decoherence}. At high temperature, the electron's thermally activated transitions are fast enough that no information about the nuclear state is revealed.

Ultimately, at high enough temperature, the MQME we use will become invalid.
The approximations used in its derivation amount to an assumption that the decay rate of the system is determined by sampling the bath spectral density at particular frequencies given by system eigenstate energy differences. A system decay rate such as $\decay_+$ corresponds to an effective linewidth of the system, and so  at high enough temperature, as $\decay_+$ increases, this linewidth will approach or even exceed the width of the spectral density. However, it is important to note that this
breakdown of the Markov approximation depends specifically on the peak width in
the density of states, and so the temperature at which this breaks down is
unconnected with the temperature required for the enhanced coherence.  We 
return to this point in more detail in the following subsection.

\subsection{Semiclassical Model}
\label{sec:semimod}

To investigate further the validity of our model in the high temperature regime we now develop a semiclassical description of the behaviour of the nuclear spin coherence. To do this, we consider the electron spin to readily lose any coherence it develops, through interaction with the high temperature bath; it can thus be effectively described as a classical two state system,  randomly fluctuating between spin-up and spin-down states. Let us denote the fluctuation rate from $\sigma_e^z=-1$ to $\sigma_e^z=1$ as $\lambda^\uparrow$ and that for the reverse process as $\lambda^\downarrow$. Assuming the rate of such fluctuations to be determined independent of the nuclear spin state at these high temperatures, we may write $\lambda^\uparrow = 2J(\omega_e)n(\omega_e)$ and $\lambda^\downarrow = 2J(\omega_e)(n(\omega_e)+1)$. Then assuming $n(\omega_e)\gg1$ for all relevant temperatures 
$\lambda^\uparrow=\lambda^\downarrow \equiv \lambda $
with 
\be
\label{eq:lamrate}
\lambda = 2J(\omega_e)n(\omega_e).
\ee

As the electron spin fluctuates, the effective energy splitting $\omega_n(t)$ of the nuclear spin flips randomly between $-2g$ (for $\sigma_e^z=-1$) and $+2g$ ($\sigma_e^z=1$), with the rate $\lambda$, and is therefore random telegraph noise (RTN), whose properties are well understood~\cite{miller}. Under this assumption, the total nuclear coherence $C_n$ may be written
\be
C_n(t) = C_n(0) \left\langle \exp\left[i\int_0^t \omega_n(s) \,ds \right]\right\rangle,
\ee
where $\langle..\rangle$ denotes an ensemble average. A cumulant expansion allows the ensemble averaged exponential to be written as an exponential function of ensemble averaged correlation functions:
\begin{widetext}
\begin{equation}
\left\langle \exp\left[ i\int_0^t \omega_n(s) \, ds \right] \right\rangle = 
\exp \left[ \sum_n \frac{i^n}{n!} \int_0^t ds_1 \int_0^t ds_2...\int_0^t ds_n \langle\langle \omega_n (s_1)\omega_n (s_2)...\omega_n (s_n) \rangle\rangle  \right]
\end{equation}
where $\langle\langle..\rangle\rangle$ indicates a cumulant. 

Our RTN has zero mean, and indeed all odd cumulants are strictly zero, and so to fourth order we may write the coherence as:
\begin{multline}
C_n(t) = C_n(0) \exp \left[  -\frac{1}{2!} \langle \omega_{n,1} \omega_{n,2} \rangle  \right.\\+
\left. \frac{1}{4!}\left(\langle \omega_{n,1} \omega_{n,2} \omega_{n,3} \omega_{n,4}\rangle-\langle \omega_{n,1} \omega_{n,2}\rangle\langle \omega_{n,3} \omega_{n,4}\rangle-\langle \omega_{n,1} \omega_{n,3}\rangle\langle \omega_{n,2} \omega_{n,4}\rangle-\langle \omega_{n,1} \omega_{n,4}\rangle\langle \omega_{n,2} \omega_{n,3}\rangle 
\right)\right]
\end{multline}
where we have explicitly written the fourth order cumulant in terms of the correlation functions (or joint moments) defined by
\be
\langle \omega_{n,1} \omega_{n,2} \rangle \equiv 2!  \int_0^t ds_2 \int_0^{s_2} ds_1 \langle \omega_n(s_1) \omega_n (s_2)\rangle ,
\ee
and
\be
\langle \omega_{n,1} \omega_{n,2} \omega_{n,3} \omega_{n,4} \rangle = 4! \int_0^t ds_4 \int_0^{s_4} ds_3\int_0^{s_3} ds_2 \int_0^{s_2} ds_1 \,\langle \omega_n(s_1) \omega_n (s_2)\omega_n(s_3) \omega_n (s_4)\rangle .
\ee
Evaluating the correlation functions for RTN we find that for the time orderings we require, we have:
\be
\langle \omega_n(s_1) \omega_n (s_2) \rangle = 4g^2\exp\left[-2\lambda(s_2-s_1)\right]
\ee
and
\be
\langle \omega_n(s_1) \omega_n (s_2)\omega_n(s_3) \omega_n (s_4) \rangle = 16g^4\exp\left[-2\lambda(s_4-s_3+s_2-s_1)\right]
\ee
where $\lambda$ is the decay rate defined above. 
\end{widetext}
Performing the integrals and assuming $\lambda t \gg 1$ yields:
\be
C_n(t) = C_n(0) \exp \left( -\frac{2g^2 t}{\lambda}\left[1  - \frac{g^2}{\lambda^2}\right]\right).
\ee
For a sufficiently fast decay $g/\lambda\ll 1$, the fourth order terms are suppressed and we find that the nuclear coherence is well approximated by exponential decay with a rate $\kappa_{\rm SC}=2g^2/\lambda$, or from Eq.~\ref{eq:lamrate}, $\kappa_{\rm SC}=g^2/J(\omega_e)n(\omega_e)$. This quantity decreases with temperature, leading to a {\it longer} exponential decay time. This is precisely the origin of the well-known motional narrowing effect in NMR~\cite{slichter1996}.
%Moreover, we find that in the high temperature limit $\kappa_{\rm SC}=\kappa_-$, and so the quantum and semiclassical models agree with each other.

\subsection{Comparison of quantum and classical results}
\label{sec:comp-quant-class}

\begin{figure}
\begin{center}
\includegraphics[width=0.5\textwidth]{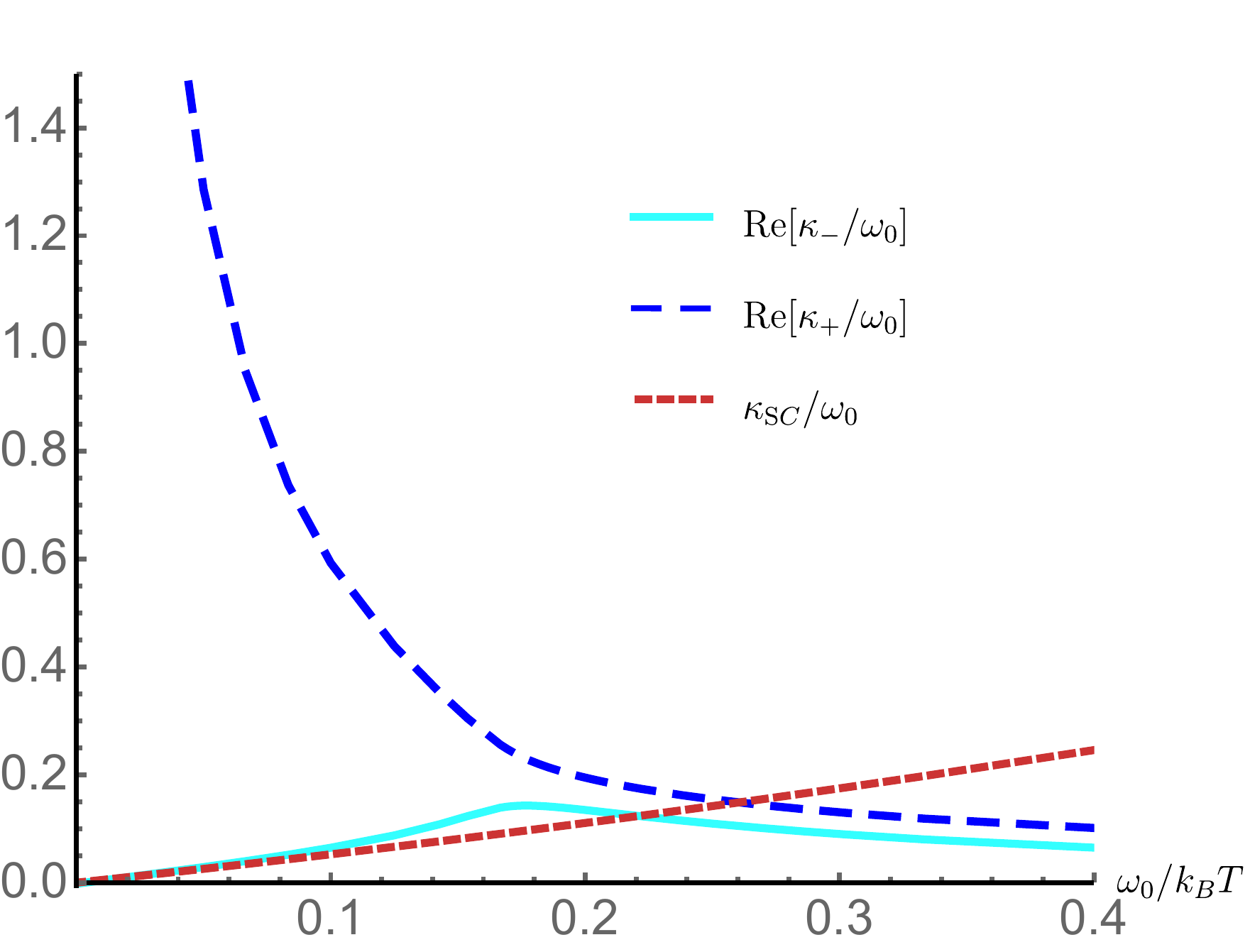}
\caption{The classical decay rate, based on the RTS model, compared with the two quantum decay rates $\kappa_+$ and $\kappa_-$ defined in Eq.~\ref{eqkappa}. Good agreement between $\kappa_-$ and the classical rate $\kappa_{\rm SC}$ can be seen in the high temperature limit, and the two theories also agree across a temperature range where the Markov approximation is still valid (see text for details).}
\label{compareclassical}
\end{center}
\end{figure}

We now calculate this classical decoherence rate as a function of temperature for our model and compare it to the quantum case discussed in Sec.~\ref{sec:MQME} above. We show the results of such a comparison in Fig.~\ref{compareclassical}. 
For the curves shown in the figure we have used a sub-Ohmic spectral density, Eq.~\ref{eq:spec} with $s=1/2$, and $\xi_c=3\omega_0$. Other parameters are the same as we have used before: $\gamma_0 = 0.02\omega_0$, $g=0.1\omega_0$. For high temperature, $\omega_0/k_BT\lesssim 0.1$, we find good agreement between the classical decay rate $\kappa_{\rm SC}$ and the slower $\kappa_-$ rate (which corresponds in the quantum model to the decay rate of the nuclear coherence following the initial electron spin decay). Indeed, at high temperature we found earlier that  $\Re[\decay_-]\simeq 4 g^2 /(\gamma^\uparrow+\gamma^\downarrow) = g^2/J(\omega_e)n(\omega_e)= \kappa_{\rm SC}$, and so we expect to see such agreement in Fig.~\ref{compareclassical}.

As noted above, to ensure that our high temperature results are robust, we should consider the validity of the Markov approximation made in deriving our MQME in this limit. 
In order to assess the region of Fig.~\ref{compareclassical} in which we can have full confidence in Markov approximation, we first find the width ${\cal W}$ for this spectral density function by calculating its second moment. This leads to ${\cal W} =  7.3\, \omega_0$. It then seems reasonable to assume that whenever the decay rates of the system are significantly smaller than ${\cal W}$ that the Markov approximation is valid. Referring to Fig.~\ref{compareclassical}, we see that the larger decay rate for the MQME, $\kappa_+$ is less that $0.2\,{\cal W}$ for $\omega_0/k_BT>0.05$. We can see therefore that the classical model and quantum model agree in a region where the Markovian approximation is valid -- i.e. for around $0.05>\omega_0/k_BT>0.1$. Both models then agree well all the way to infinite temperature, and so the conclusion that we made above - that heating a system can improve its coherence time - is indeed valid.

\section{Discussion and Conclusion}
\label{discussion}

In order to observe the protected coherences that we have described, it is necessary to engineer a system whereby the transition spacings are comparable to their linewidths, and which can be tuned. Quantum dots (QDs) doped with a single electron have spin selective optical transitions to trion states, so forming a hybrid spin-photon system. Sweeney {\it et al.} \cite{Sweeney2014} found that the optical transition spacings in a cavity-QD system were $2.8$ times greater than the linewidths. Purcell enhancement in QD systems can broaden linewidths by a factor of $6.7$ \cite{Weiler2011} to produce highly indistinguishable photons, which would allow coherences to survive even at zero temperature.

Closely spaced transitions are also found in NV centres in diamond, already the focus of many quantum computing studies. The NV centre carries an electron spin which can be optically excited via spin selective transitions \cite{Bourgeois2015}.
The spin sublevel splitting can be tuned in zero magnetic field via an externally applied electric field \cite{Bourgeois2015}.
By placing the diamond inside a tunable cavity, the emission rate of the NV centre can also be enhanced, increasing transition linewidths \cite{Johnson2015} -- thus providing enough flexibility to observe coherence preservation during decay.

We have shown that decoherence of a system does not apply equally to all its parts, and that a combination of the correct coupling and temperature tuning can preserve the coherence of part of a quantum system. These effects could be exploited in the next generation of hybrid quantum information processing devices.

%We thank T. M. Stace for useful discussions. 

HMC acknowledges studentship funding from EPSRC under grant no. EP/G03673X/1. PGK acknowledges support
  from EPSRC (EP/M010910/1). BWL acknowledges support from EPSRC (EP/K025562/1).
PRE acknowledges funding from SFI (15/IACA/3402).
JK acknowledges financial support from EPSRC programs ``TOPNES'' (EP/I031014/1)
and ``Hybrid-Polaritonics'' (EP/M025330/1).

\bibliographystyle{apsrev4-1}

\begin{thebibliography}{29}%
\makeatletter
\providecommand \@ifxundefined [1]{%
 \@ifx{#1\undefined}
}%
\providecommand \@ifnum [1]{%
 \ifnum #1\expandafter \@firstoftwo
 \else \expandafter \@secondoftwo
 \fi
}%
\providecommand \@ifx [1]{%
 \ifx #1\expandafter \@firstoftwo
 \else \expandafter \@secondoftwo
 \fi
}%
\providecommand \natexlab [1]{#1}%
\providecommand \enquote  [1]{``#1''}%
\providecommand \bibnamefont  [1]{#1}%
\providecommand \bibfnamefont [1]{#1}%
\providecommand \citenamefont [1]{#1}%
\providecommand \href@noop [0]{\@secondoftwo}%
\providecommand \href [0]{\begingroup \@sanitize@url \@href}%
\providecommand \@href[1]{\@@startlink{#1}\@@href}%
\providecommand \@@href[1]{\endgroup#1\@@endlink}%
\providecommand \@sanitize@url [0]{\catcode `\\12\catcode `\$12\catcode
  `\&12\catcode `\#12\catcode `\^12\catcode `\_12\catcode `\%12\relax}%
\providecommand \@@startlink[1]{}%
\providecommand \@@endlink[0]{}%
\providecommand \url  [0]{\begingroup\@sanitize@url \@url }%
\providecommand \@url [1]{\endgroup\@href {#1}{\urlprefix }}%
\providecommand \urlprefix  [0]{URL }%
\providecommand \Eprint [0]{\href }%
\providecommand \doibase [0]{http://dx.doi.org/}%
\providecommand \selectlanguage [0]{\@gobble}%
\providecommand \bibinfo  [0]{\@secondoftwo}%
\providecommand \bibfield  [0]{\@secondoftwo}%
\providecommand \translation [1]{[#1]}%
\providecommand \BibitemOpen [0]{}%
\providecommand \bibitemStop [0]{}%
\providecommand \bibitemNoStop [0]{.\EOS\space}%
\providecommand \EOS [0]{\spacefactor3000\relax}%
\providecommand \BibitemShut  [1]{\csname bibitem#1\endcsname}%
\let\auto@bib@innerbib\@empty
%</preamble>
\bibitem [{\citenamefont {Palma}\ \emph {et~al.}(1996)\citenamefont {Palma},
  \citenamefont {Suominen},\ and\ \citenamefont {Ekert}}]{ekert97}%
  \BibitemOpen
  \bibfield  {author} {\bibinfo {author} {\bibfnamefont {G.~M.}\ \bibnamefont
  {Palma}}, \bibinfo {author} {\bibfnamefont {K.-A.}\ \bibnamefont {Suominen}},
  \ and\ \bibinfo {author} {\bibfnamefont {A.~K.}\ \bibnamefont {Ekert}},\
  }\href {\doibase 10.1098/rspa.1996.0029} {\bibfield  {journal} {\bibinfo
  {journal} {Proc. Roy. Soc. Lond. A}\ }\textbf {\bibinfo {volume} {452}},\
  \bibinfo {pages} {567} (\bibinfo {year} {1996})}\BibitemShut {NoStop}%
\bibitem [{\citenamefont {Breuer}\ and\ \citenamefont
  {Petruccione}(2002)}]{breuerpet}%
  \BibitemOpen
  \bibfield  {author} {\bibinfo {author} {\bibfnamefont {H.}~\bibnamefont
  {Breuer}}\ and\ \bibinfo {author} {\bibfnamefont {F.}~\bibnamefont
  {Petruccione}},\ }\href@noop {} {\emph {\bibinfo {title} {{The Theory of Open
  Quantum Systems}}}}\ (\bibinfo  {publisher} {Oxford University Press},\
  \bibinfo {year} {2002})\BibitemShut {NoStop}%
\bibitem [{\citenamefont {Schlosshauer}(2005)}]{schlosshauer05}%
  \BibitemOpen
  \bibfield  {author} {\bibinfo {author} {\bibfnamefont {M.}~\bibnamefont
  {Schlosshauer}},\ }\href {\doibase 10.1103/RevModPhys.76.1267} {\bibfield
  {journal} {\bibinfo  {journal} {Rev. Mod. Phys.}\ }\textbf {\bibinfo {volume}
  {76}},\ \bibinfo {pages} {1267} (\bibinfo {year} {2005})}\BibitemShut
  {NoStop}%
\bibitem [{\citenamefont {Paz}\ and\ \citenamefont {Zurek}(1999)}]{paz99}%
  \BibitemOpen
  \bibfield  {author} {\bibinfo {author} {\bibfnamefont {J.~P.}\ \bibnamefont
  {Paz}}\ and\ \bibinfo {author} {\bibfnamefont {W.~H.}\ \bibnamefont
  {Zurek}},\ }\href {\doibase 10.1103/PhysRevLett.82.5181} {\bibfield
  {journal} {\bibinfo  {journal} {Phys. Rev. Lett.}\ }\textbf {\bibinfo
  {volume} {82}},\ \bibinfo {pages} {5181} (\bibinfo {year}
  {1999})}\BibitemShut {NoStop}%
\bibitem [{\citenamefont {Zurek}(2003)}]{zurek03}%
  \BibitemOpen
  \bibfield  {author} {\bibinfo {author} {\bibfnamefont {W.~H.}\ \bibnamefont
  {Zurek}},\ }\href {\doibase 10.1103/RevModPhys.75.715} {\bibfield  {journal}
  {\bibinfo  {journal} {Rev. Mod. Phys.}\ }\textbf {\bibinfo {volume} {75}},\
  \bibinfo {pages} {715} (\bibinfo {year} {2003})}\BibitemShut {NoStop}%
\bibitem [{\citenamefont {Atat\"{u}re}\ \emph {et~al.}(2007)\citenamefont
  {Atat\"{u}re}, \citenamefont {Dreiser}, \citenamefont {Badolato},\ and\
  \citenamefont {Imamoglu}}]{atature07}%
  \BibitemOpen
  \bibfield  {author} {\bibinfo {author} {\bibfnamefont {M.}~\bibnamefont
  {Atat\"{u}re}}, \bibinfo {author} {\bibfnamefont {J.}~\bibnamefont
  {Dreiser}}, \bibinfo {author} {\bibfnamefont {A.}~\bibnamefont {Badolato}}, \
  and\ \bibinfo {author} {\bibfnamefont {A.}~\bibnamefont {Imamoglu}},\ }\href
  {http://dx.doi.org/10.1038/nphys521} {\bibfield  {journal} {\bibinfo
  {journal} {Nat. Phys.}\ }\textbf {\bibinfo {volume} {3}},\ \bibinfo {pages}
  {101} (\bibinfo {year} {2007})}\BibitemShut {NoStop}%
\bibitem [{\citenamefont {Ramsay}\ \emph {et~al.}(2008)\citenamefont {Ramsay},
  \citenamefont {Boyle}, \citenamefont {Kolodka}, \citenamefont {Oliveira},
  \citenamefont {Skiba-Szymanska}, \citenamefont {Liu}, \citenamefont
  {Hopkinson}, \citenamefont {Fox},\ and\ \citenamefont {Skolnick}}]{ramsay08}%
  \BibitemOpen
  \bibfield  {author} {\bibinfo {author} {\bibfnamefont {A.~J.}\ \bibnamefont
  {Ramsay}}, \bibinfo {author} {\bibfnamefont {S.~J.}\ \bibnamefont {Boyle}},
  \bibinfo {author} {\bibfnamefont {R.~S.}\ \bibnamefont {Kolodka}}, \bibinfo
  {author} {\bibfnamefont {J.~B.~B.}\ \bibnamefont {Oliveira}}, \bibinfo
  {author} {\bibfnamefont {J.}~\bibnamefont {Skiba-Szymanska}}, \bibinfo
  {author} {\bibfnamefont {H.~Y.}\ \bibnamefont {Liu}}, \bibinfo {author}
  {\bibfnamefont {M.}~\bibnamefont {Hopkinson}}, \bibinfo {author}
  {\bibfnamefont {A.~M.}\ \bibnamefont {Fox}}, \ and\ \bibinfo {author}
  {\bibfnamefont {M.~S.}\ \bibnamefont {Skolnick}},\ }\href {\doibase
  10.1103/PhysRevLett.100.197401} {\bibfield  {journal} {\bibinfo  {journal}
  {Phys. Rev. Lett.}\ }\textbf {\bibinfo {volume} {100}},\ \bibinfo {pages}
  {197401} (\bibinfo {year} {2008})}\BibitemShut {NoStop}%
\bibitem [{\citenamefont {Schaffry}\ \emph {et~al.}(2010)\citenamefont
  {Schaffry}, \citenamefont {Filidou}, \citenamefont {Karlen}, \citenamefont
  {Gauger}, \citenamefont {Benjamin}, \citenamefont {Anderson}, \citenamefont
  {Ardavan}, \citenamefont {Briggs}, \citenamefont {Maeda}, \citenamefont
  {Henbest}, \citenamefont {Giustino}, \citenamefont {Morton},\ and\
  \citenamefont {Lovett}}]{schaffry10}%
  \BibitemOpen
  \bibfield  {author} {\bibinfo {author} {\bibfnamefont {M.}~\bibnamefont
  {Schaffry}}, \bibinfo {author} {\bibfnamefont {V.}~\bibnamefont {Filidou}},
  \bibinfo {author} {\bibfnamefont {S.~D.}\ \bibnamefont {Karlen}}, \bibinfo
  {author} {\bibfnamefont {E.~M.}\ \bibnamefont {Gauger}}, \bibinfo {author}
  {\bibfnamefont {S.~C.}\ \bibnamefont {Benjamin}}, \bibinfo {author}
  {\bibfnamefont {H.~L.}\ \bibnamefont {Anderson}}, \bibinfo {author}
  {\bibfnamefont {A.}~\bibnamefont {Ardavan}}, \bibinfo {author} {\bibfnamefont
  {G.~A.~D.}\ \bibnamefont {Briggs}}, \bibinfo {author} {\bibfnamefont
  {K.}~\bibnamefont {Maeda}}, \bibinfo {author} {\bibfnamefont {K.~B.}\
  \bibnamefont {Henbest}}, \bibinfo {author} {\bibfnamefont {F.}~\bibnamefont
  {Giustino}}, \bibinfo {author} {\bibfnamefont {J.~J.~L.}\ \bibnamefont
  {Morton}}, \ and\ \bibinfo {author} {\bibfnamefont {B.~W.}\ \bibnamefont
  {Lovett}},\ }\href {\doibase 10.1103/PhysRevLett.104.200501} {\bibfield
  {journal} {\bibinfo  {journal} {Phys. Rev. Lett.}\ }\textbf {\bibinfo
  {volume} {104}},\ \bibinfo {pages} {200501} (\bibinfo {year}
  {2010})}\BibitemShut {NoStop}%
\bibitem [{\citenamefont {Neumann}\ \emph {et~al.}(2010)\citenamefont
  {Neumann}, \citenamefont {Beck}, \citenamefont {Steiner}, \citenamefont
  {Rempp}, \citenamefont {Fedder}, \citenamefont {Hemmer}, \citenamefont
  {Wrachtrup},\ and\ \citenamefont {Jelezko}}]{neumann10}%
  \BibitemOpen
  \bibfield  {author} {\bibinfo {author} {\bibfnamefont {P.}~\bibnamefont
  {Neumann}}, \bibinfo {author} {\bibfnamefont {J.}~\bibnamefont {Beck}},
  \bibinfo {author} {\bibfnamefont {M.}~\bibnamefont {Steiner}}, \bibinfo
  {author} {\bibfnamefont {F.}~\bibnamefont {Rempp}}, \bibinfo {author}
  {\bibfnamefont {H.}~\bibnamefont {Fedder}}, \bibinfo {author} {\bibfnamefont
  {P.~R.}\ \bibnamefont {Hemmer}}, \bibinfo {author} {\bibfnamefont
  {J.}~\bibnamefont {Wrachtrup}}, \ and\ \bibinfo {author} {\bibfnamefont
  {F.}~\bibnamefont {Jelezko}},\ }\href {\doibase 10.1126/science.1189075}
  {\bibfield  {journal} {\bibinfo  {journal} {Science}\ }\textbf {\bibinfo
  {volume} {329}},\ \bibinfo {pages} {542} (\bibinfo {year}
  {2010})}\BibitemShut {NoStop}%
\bibitem [{\citenamefont {Xiang}\ \emph {et~al.}(2013)\citenamefont {Xiang},
  \citenamefont {Ashhab}, \citenamefont {You},\ and\ \citenamefont
  {Nori}}]{nori13}%
  \BibitemOpen
  \bibfield  {author} {\bibinfo {author} {\bibfnamefont {Z.-L.}\ \bibnamefont
  {Xiang}}, \bibinfo {author} {\bibfnamefont {S.}~\bibnamefont {Ashhab}},
  \bibinfo {author} {\bibfnamefont {J.~Q.}\ \bibnamefont {You}}, \ and\
  \bibinfo {author} {\bibfnamefont {F.}~\bibnamefont {Nori}},\ }\href {\doibase
  10.1103/RevModPhys.85.623} {\bibfield  {journal} {\bibinfo  {journal} {Rev.
  Mod. Phys.}\ }\textbf {\bibinfo {volume} {85}},\ \bibinfo {pages} {623}
  (\bibinfo {year} {2013})}\BibitemShut {NoStop}%
\bibitem [{\citenamefont {Marcos}\ \emph {et~al.}(2010)\citenamefont {Marcos},
  \citenamefont {Wubs}, \citenamefont {Taylor}, \citenamefont {Aguado},
  \citenamefont {Lukin},\ and\ \citenamefont {S\o{}rensen}}]{marcos10}%
  \BibitemOpen
  \bibfield  {author} {\bibinfo {author} {\bibfnamefont {D.}~\bibnamefont
  {Marcos}}, \bibinfo {author} {\bibfnamefont {M.}~\bibnamefont {Wubs}},
  \bibinfo {author} {\bibfnamefont {J.~M.}\ \bibnamefont {Taylor}}, \bibinfo
  {author} {\bibfnamefont {R.}~\bibnamefont {Aguado}}, \bibinfo {author}
  {\bibfnamefont {M.~D.}\ \bibnamefont {Lukin}}, \ and\ \bibinfo {author}
  {\bibfnamefont {A.~S.}\ \bibnamefont {S\o{}rensen}},\ }\href {\doibase
  10.1103/PhysRevLett.105.210501} {\bibfield  {journal} {\bibinfo  {journal}
  {Phys. Rev. Lett.}\ }\textbf {\bibinfo {volume} {105}},\ \bibinfo {pages}
  {210501} (\bibinfo {year} {2010})}\BibitemShut {NoStop}%
\bibitem [{\citenamefont {Morton}\ and\ \citenamefont
  {Lovett}(2011)}]{Morton2011}%
  \BibitemOpen
  \bibfield  {author} {\bibinfo {author} {\bibfnamefont {J.~J.~L.}\
  \bibnamefont {Morton}}\ and\ \bibinfo {author} {\bibfnamefont {B.~W.}\
  \bibnamefont {Lovett}},\ }\href {\doibase
  10.1146/annurev-conmatphys-062910-140514} {\bibfield  {journal} {\bibinfo
  {journal} {Ann. Rev. Cond. Matter Phys.}\ }\textbf {\bibinfo {volume} {2}},\
  \bibinfo {pages} {189} (\bibinfo {year} {2011})}\BibitemShut {NoStop}%
\bibitem [{\citenamefont {Kurizki}\ \emph {et~al.}(2015)\citenamefont
  {Kurizki}, \citenamefont {Bertet}, \citenamefont {Kubo}, \citenamefont
  {M\o{}lmer}, \citenamefont {Petrosyan}, \citenamefont {Rabl},\ and\
  \citenamefont {Schmiedmayer}}]{Kurizki2015}%
  \BibitemOpen
  \bibfield  {author} {\bibinfo {author} {\bibfnamefont {G.}~\bibnamefont
  {Kurizki}}, \bibinfo {author} {\bibfnamefont {P.}~\bibnamefont {Bertet}},
  \bibinfo {author} {\bibfnamefont {Y.}~\bibnamefont {Kubo}}, \bibinfo {author}
  {\bibfnamefont {K.}~\bibnamefont {M\o{}lmer}}, \bibinfo {author}
  {\bibfnamefont {D.}~\bibnamefont {Petrosyan}}, \bibinfo {author}
  {\bibfnamefont {P.}~\bibnamefont {Rabl}}, \ and\ \bibinfo {author}
  {\bibfnamefont {J.}~\bibnamefont {Schmiedmayer}},\ }\href {\doibase
  10.1073/pnas.1419326112} {\bibfield  {journal} {\bibinfo  {journal} {Proc.
  Natl. Acad. Sci.}\ }\textbf {\bibinfo {volume} {112}},\ \bibinfo {pages}
  {201419326} (\bibinfo {year} {2015})}\BibitemShut {NoStop}%
\bibitem [{\citenamefont {Morley}\ \emph {et~al.}(2012)\citenamefont {Morley},
  \citenamefont {Lueders}, \citenamefont {{Hamed Mohammady}}, \citenamefont
  {Balian}, \citenamefont {Aeppli}, \citenamefont {Kay}, \citenamefont
  {Witzel}, \citenamefont {Jeschke},\ and\ \citenamefont
  {Monteiro}}]{Morley2012}%
  \BibitemOpen
  \bibfield  {author} {\bibinfo {author} {\bibfnamefont {G.~W.}\ \bibnamefont
  {Morley}}, \bibinfo {author} {\bibfnamefont {P.}~\bibnamefont {Lueders}},
  \bibinfo {author} {\bibfnamefont {M.}~\bibnamefont {{Hamed Mohammady}}},
  \bibinfo {author} {\bibfnamefont {S.~J.}\ \bibnamefont {Balian}}, \bibinfo
  {author} {\bibfnamefont {G.}~\bibnamefont {Aeppli}}, \bibinfo {author}
  {\bibfnamefont {C.~W.~M.}\ \bibnamefont {Kay}}, \bibinfo {author}
  {\bibfnamefont {W.~M.}\ \bibnamefont {Witzel}}, \bibinfo {author}
  {\bibfnamefont {G.}~\bibnamefont {Jeschke}}, \ and\ \bibinfo {author}
  {\bibfnamefont {T.~S.}\ \bibnamefont {Monteiro}},\ }\href {\doibase
  10.1038/nmat3499} {\bibfield  {journal} {\bibinfo  {journal} {Nat. Mater.}\
  }\textbf {\bibinfo {volume} {12}},\ \bibinfo {pages} {103} (\bibinfo {year}
  {2012})}\BibitemShut {NoStop}%
\bibitem [{\citenamefont {Zhao}\ and\ \citenamefont
  {Wrachtrup}(2013)}]{Zhao2013}%
  \BibitemOpen
  \bibfield  {author} {\bibinfo {author} {\bibfnamefont {N.}~\bibnamefont
  {Zhao}}\ and\ \bibinfo {author} {\bibfnamefont {J.}~\bibnamefont
  {Wrachtrup}},\ }\href {\doibase 10.1038/nmat3531} {\bibfield  {journal}
  {\bibinfo  {journal} {Nat. Mater.}\ }\textbf {\bibinfo {volume} {12}},\
  \bibinfo {pages} {97} (\bibinfo {year} {2013})}\BibitemShut {NoStop}%
\bibitem [{\citenamefont {Zhong}\ \emph {et~al.}(2015)\citenamefont {Zhong},
  \citenamefont {Hedges}, \citenamefont {Ahlefeldt}, \citenamefont
  {Bartholomew}, \citenamefont {Beavan}, \citenamefont {Wittig}, \citenamefont
  {Longdell},\ and\ \citenamefont {Sellars}}]{Zhong2015}%
  \BibitemOpen
  \bibfield  {author} {\bibinfo {author} {\bibfnamefont {M.}~\bibnamefont
  {Zhong}}, \bibinfo {author} {\bibfnamefont {M.~P.}\ \bibnamefont {Hedges}},
  \bibinfo {author} {\bibfnamefont {R.~L.}\ \bibnamefont {Ahlefeldt}}, \bibinfo
  {author} {\bibfnamefont {J.~G.}\ \bibnamefont {Bartholomew}}, \bibinfo
  {author} {\bibfnamefont {S.~E.}\ \bibnamefont {Beavan}}, \bibinfo {author}
  {\bibfnamefont {S.~M.}\ \bibnamefont {Wittig}}, \bibinfo {author}
  {\bibfnamefont {J.~J.}\ \bibnamefont {Longdell}}, \ and\ \bibinfo {author}
  {\bibfnamefont {M.~J.}\ \bibnamefont {Sellars}},\ }\href {\doibase
  10.1038/nature14025} {\bibfield  {journal} {\bibinfo  {journal} {Nature}\
  }\textbf {\bibinfo {volume} {517}},\ \bibinfo {pages} {177} (\bibinfo {year}
  {2015})}\BibitemShut {NoStop}%
\bibitem [{\citenamefont {Bienfait}\ \emph {et~al.}(2016)\citenamefont
  {Bienfait}, \citenamefont {Pla}, \citenamefont {Kubo}, \citenamefont {Zhou},
  \citenamefont {Stern}, \citenamefont {Lo}, \citenamefont {Weis},
  \citenamefont {Schenkel}, \citenamefont {Vion}, \citenamefont {Esteve},
  \citenamefont {Morton},\ and\ \citenamefont {Bertet}}]{bertet16}%
  \BibitemOpen
  \bibfield  {author} {\bibinfo {author} {\bibfnamefont {A.}~\bibnamefont
  {Bienfait}}, \bibinfo {author} {\bibfnamefont {J.~J.}\ \bibnamefont {Pla}},
  \bibinfo {author} {\bibfnamefont {Y.}~\bibnamefont {Kubo}}, \bibinfo {author}
  {\bibfnamefont {X.}~\bibnamefont {Zhou}}, \bibinfo {author} {\bibfnamefont
  {M.}~\bibnamefont {Stern}}, \bibinfo {author} {\bibfnamefont {C.~C.}\
  \bibnamefont {Lo}}, \bibinfo {author} {\bibfnamefont {C.~D.}\ \bibnamefont
  {Weis}}, \bibinfo {author} {\bibfnamefont {T.}~\bibnamefont {Schenkel}},
  \bibinfo {author} {\bibfnamefont {D.}~\bibnamefont {Vion}}, \bibinfo {author}
  {\bibfnamefont {D.}~\bibnamefont {Esteve}}, \bibinfo {author} {\bibfnamefont
  {J.~J.~L.}\ \bibnamefont {Morton}}, \ and\ \bibinfo {author} {\bibfnamefont
  {P.}~\bibnamefont {Bertet}},\ }\href {http://dx.doi.org/10.1038/nature16944}
  {\bibfield  {journal} {\bibinfo  {journal} {Nature}\ }\textbf {\bibinfo
  {volume} {531}},\ \bibinfo {pages} {74} (\bibinfo {year} {2016})}\BibitemShut
  {NoStop}%
\bibitem [{\citenamefont {Weisskopf}\ and\ \citenamefont
  {Wigner}(1930)}]{Wigner1930}%
  \BibitemOpen
  \bibfield  {author} {\bibinfo {author} {\bibfnamefont {V.}~\bibnamefont
  {Weisskopf}}\ and\ \bibinfo {author} {\bibfnamefont {E.}~\bibnamefont
  {Wigner}},\ }\href {\doibase 10.1007/BF01336768} {\bibfield  {journal}
  {\bibinfo  {journal} {Zeitschrift f{\"u}r Physik}\ }\textbf {\bibinfo
  {volume} {63}},\ \bibinfo {pages} {54} (\bibinfo {year} {1930})}\BibitemShut
  {NoStop}%
\bibitem [{\citenamefont {Vacchini}\ and\ \citenamefont
  {Breuer}(2010)}]{Vacchini2010}%
  \BibitemOpen
  \bibfield  {author} {\bibinfo {author} {\bibfnamefont {B.}~\bibnamefont
  {Vacchini}}\ and\ \bibinfo {author} {\bibfnamefont {H.-P.}\ \bibnamefont
  {Breuer}},\ }\href {\doibase 10.1103/PhysRevA.81.042103} {\bibfield
  {journal} {\bibinfo  {journal} {Phys. Rev. A}\ }\textbf {\bibinfo {volume}
  {81}},\ \bibinfo {pages} {042103} (\bibinfo {year} {2010})}\BibitemShut
  {NoStop}%
\bibitem [{\citenamefont {Stace}\ \emph {et~al.}(2013)\citenamefont {Stace},
  \citenamefont {Doherty},\ and\ \citenamefont
  {Reilly}}]{PhysRevLett.111.180602}%
  \BibitemOpen
  \bibfield  {author} {\bibinfo {author} {\bibfnamefont {T.~M.}\ \bibnamefont
  {Stace}}, \bibinfo {author} {\bibfnamefont {A.~C.}\ \bibnamefont {Doherty}},
  \ and\ \bibinfo {author} {\bibfnamefont {D.~J.}\ \bibnamefont {Reilly}},\
  }\href {\doibase 10.1103/PhysRevLett.111.180602} {\bibfield  {journal}
  {\bibinfo  {journal} {Phys. Rev. Lett.}\ }\textbf {\bibinfo {volume} {111}},\
  \bibinfo {pages} {180602} (\bibinfo {year} {2013})}\BibitemShut {NoStop}%
\bibitem [{\citenamefont {Eastham}\ \emph {et~al.}(2016)\citenamefont
  {Eastham}, \citenamefont {Kirton}, \citenamefont {Cammack}, \citenamefont
  {Lovett},\ and\ \citenamefont {Keeling}}]{Eastham16}%
  \BibitemOpen
  \bibfield  {author} {\bibinfo {author} {\bibfnamefont {P.~R.}\ \bibnamefont
  {Eastham}}, \bibinfo {author} {\bibfnamefont {P.}~\bibnamefont {Kirton}},
  \bibinfo {author} {\bibfnamefont {H.~M.}\ \bibnamefont {Cammack}}, \bibinfo
  {author} {\bibfnamefont {B.~W.}\ \bibnamefont {Lovett}}, \ and\ \bibinfo
  {author} {\bibfnamefont {J.}~\bibnamefont {Keeling}},\ }\href {\doibase
  10.1103/PhysRevA.94.012110} {\bibfield  {journal} {\bibinfo  {journal} {Phys.
  Rev. A}\ }\textbf {\bibinfo {volume} {94}},\ \bibinfo {pages} {012110}
  (\bibinfo {year} {2016})}\BibitemShut {NoStop}%
\bibitem [{\citenamefont {Lindblad}(1976)}]{Lindblad1976}%
  \BibitemOpen
  \bibfield  {author} {\bibinfo {author} {\bibfnamefont {G.}~\bibnamefont
  {Lindblad}},\ }\href@noop {} {\bibfield  {journal} {\bibinfo  {journal}
  {Comm. Math. Phys.}\ }\textbf {\bibinfo {volume} {48}},\ \bibinfo {pages}
  {119} (\bibinfo {year} {1976})}\BibitemShut {NoStop}%
\bibitem [{\citenamefont {D{\"{u}}mcke}\ and\ \citenamefont
  {Spohn}(1979)}]{Dumcke1979}%
  \BibitemOpen
  \bibfield  {author} {\bibinfo {author} {\bibfnamefont {R.}~\bibnamefont
  {D{\"{u}}mcke}}\ and\ \bibinfo {author} {\bibfnamefont {H.}~\bibnamefont
  {Spohn}},\ }\href {http://www.springerlink.com/index/v708585804823525.pdf}
  {\bibfield  {journal} {\bibinfo  {journal} {Z. Phys. B}\ }\textbf {\bibinfo
  {volume} {34}},\ \bibinfo {pages} {419} (\bibinfo {year} {1979})}\BibitemShut
  {NoStop}%
\bibitem [{\citenamefont {Jeske}\ \emph
  {et~al.}(2015{\natexlab{a}})\citenamefont {Jeske}, \citenamefont {Ing},
  \citenamefont {Plenio}, \citenamefont {Huelga},\ and\ \citenamefont
  {Cole}}]{Jeske2015}%
  \BibitemOpen
  \bibfield  {author} {\bibinfo {author} {\bibfnamefont {J.}~\bibnamefont
  {Jeske}}, \bibinfo {author} {\bibfnamefont {D.~J.}\ \bibnamefont {Ing}},
  \bibinfo {author} {\bibfnamefont {M.~B.}\ \bibnamefont {Plenio}}, \bibinfo
  {author} {\bibfnamefont {S.~F.}\ \bibnamefont {Huelga}}, \ and\ \bibinfo
  {author} {\bibfnamefont {J.~H.}\ \bibnamefont {Cole}},\ }\href {\doibase
  10.1063/1.4907370} {\bibfield  {journal} {\bibinfo  {journal} {J. Chem.
  Phys.}\ }\textbf {\bibinfo {volume} {142}},\ \bibinfo {eid} {064104}
  (\bibinfo {year} {2015}{\natexlab{a}})}\BibitemShut {NoStop}%
\bibitem [{\citenamefont {Pearle}(2012)}]{Pearle2012}%
  \BibitemOpen
  \bibfield  {author} {\bibinfo {author} {\bibfnamefont {P.}~\bibnamefont
  {Pearle}},\ }\href {\doibase 10.1088/0143-0807/33/4/805} {\bibfield
  {journal} {\bibinfo  {journal} {Europ. J. Phys.}\ }\textbf {\bibinfo {volume}
  {33}},\ \bibinfo {pages} {805} (\bibinfo {year} {2012})},\ \Eprint
  {http://arxiv.org/abs/1204.2016} {arXiv:1204.2016} \BibitemShut {NoStop}%
\bibitem [{\citenamefont {Jeske}\ \emph
  {et~al.}(2015{\natexlab{b}})\citenamefont {Jeske}, \citenamefont {Ing},
  \citenamefont {Plenio}, \citenamefont {Huelga},\ and\ \citenamefont
  {Cole}}]{Jeske2014a}%
  \BibitemOpen
  \bibfield  {author} {\bibinfo {author} {\bibfnamefont {J.}~\bibnamefont
  {Jeske}}, \bibinfo {author} {\bibfnamefont {D.}~\bibnamefont {Ing}}, \bibinfo
  {author} {\bibfnamefont {M.~B.}\ \bibnamefont {Plenio}}, \bibinfo {author}
  {\bibfnamefont {S.~F.}\ \bibnamefont {Huelga}}, \ and\ \bibinfo {author}
  {\bibfnamefont {J.~H.}\ \bibnamefont {Cole}},\ }\href {\doibase
  10.1063/1.4907370} {\bibfield  {journal} {\bibinfo  {journal} {J. Chem.
  Phys}\ }\textbf {\bibinfo {volume} {142}},\ \bibinfo {pages} {064104}
  (\bibinfo {year} {2015}{\natexlab{b}})}\BibitemShut {NoStop}%
  \bibitem [{\citenamefont {Miller}\ and\ \citenamefont
  {Childers}(2012)}]{miller}%
  \BibitemOpen
  \bibfield  {author} {\bibinfo {author} {\bibfnamefont {S.~L.}\ \bibnamefont
  {Miller}}\ and\ \bibinfo {author} {\bibfnamefont {D.}~\bibnamefont
  {Childers}},\ }in\ \href {\doibase
  https://doi.org/10.1016/B978-0-12-386981-4.50011-4} {\emph {\bibinfo
  {booktitle} {Probability and Random Processes (Second Edition)}}},\ \bibinfo
  {editor} {edited by\ \bibinfo {editor} {\bibfnamefont {S.~L.}\ \bibnamefont
  {Miller}}\ and\ \bibinfo {editor} {\bibfnamefont {D.}~\bibnamefont
  {Childers}}}\ (\bibinfo  {publisher} {Academic Press},\ \bibinfo {address}
  {Boston},\ \bibinfo {year} {2012})\ \bibinfo {edition} \
  \ pp.\ \bibinfo {pages} {335 -- 382}\BibitemShut {NoStop}%
 \bibitem [{\citenamefont {Slichter}(1996)}]{slichter1996}%
  \BibitemOpen
  \bibfield  {author} {\bibinfo {author} {\bibfnamefont {C.~P.}\ \bibnamefont
  {Slichter}},\ }\href@noop {} {\emph {\bibinfo {title} {Principles of Magnetic
  Resonance}}},\ \bibinfo {edition} {3rd}\ ed.\ (\bibinfo  {publisher}
  {Springer},\ \bibinfo {address} {Berlin},\ \bibinfo {year}
  {1996})\BibitemShut {NoStop}
\bibitem [{\citenamefont {Sweeney}\ \emph {et~al.}(2014)\citenamefont
  {Sweeney}, \citenamefont {Carter}, \citenamefont {Bracker}, \citenamefont
  {Kim}, \citenamefont {Kim}, \citenamefont {Yang}, \citenamefont {Vora},
  \citenamefont {Brereton}, \citenamefont {Cleveland},\ and\ \citenamefont
  {Gammon}}]{Sweeney2014}%
  \BibitemOpen
  \bibfield  {author} {\bibinfo {author} {\bibfnamefont {T.~M.}\ \bibnamefont
  {Sweeney}}, \bibinfo {author} {\bibfnamefont {S.~G.}\ \bibnamefont {Carter}},
  \bibinfo {author} {\bibfnamefont {A.~S.}\ \bibnamefont {Bracker}}, \bibinfo
  {author} {\bibfnamefont {M.}~\bibnamefont {Kim}}, \bibinfo {author}
  {\bibfnamefont {C.~S.}\ \bibnamefont {Kim}}, \bibinfo {author} {\bibfnamefont
  {L.}~\bibnamefont {Yang}}, \bibinfo {author} {\bibfnamefont {P.~M.}\
  \bibnamefont {Vora}}, \bibinfo {author} {\bibfnamefont {P.~G.}\ \bibnamefont
  {Brereton}}, \bibinfo {author} {\bibfnamefont {E.~R.}\ \bibnamefont
  {Cleveland}}, \ and\ \bibinfo {author} {\bibfnamefont {D.}~\bibnamefont
  {Gammon}},\ }\href {\doibase 10.1038/nphoton.2014.84} {\bibfield  {journal}
  {\bibinfo  {journal} {Nat. Photon.}\ }\textbf {\bibinfo {volume} {8}},\
  \bibinfo {pages} {442} (\bibinfo {year} {2014})},\ \Eprint
  {http://arxiv.org/abs/1402.4494} {arXiv:1402.4494} \BibitemShut {NoStop}%
\bibitem [{\citenamefont {Weiler}\ \emph {et~al.}(2011)\citenamefont {Weiler},
  \citenamefont {Ulhaq}, \citenamefont {Ulrich}, \citenamefont {Reitzenstein},
  \citenamefont {L{\"{o}}ffler}, \citenamefont {Forchel},\ and\ \citenamefont
  {Michler}}]{Weiler2011}%
  \BibitemOpen
  \bibfield  {author} {\bibinfo {author} {\bibfnamefont {S.}~\bibnamefont
  {Weiler}}, \bibinfo {author} {\bibfnamefont {A.}~\bibnamefont {Ulhaq}},
  \bibinfo {author} {\bibfnamefont {S.~M.}\ \bibnamefont {Ulrich}}, \bibinfo
  {author} {\bibfnamefont {S.}~\bibnamefont {Reitzenstein}}, \bibinfo {author}
  {\bibfnamefont {A.}~\bibnamefont {L{\"{o}}ffler}}, \bibinfo {author}
  {\bibfnamefont {A.}~\bibnamefont {Forchel}}, \ and\ \bibinfo {author}
  {\bibfnamefont {P.}~\bibnamefont {Michler}},\ }\href {\doibase
  10.1002/pssb.201000781} {\bibfield  {journal} {\bibinfo  {journal} {Phys.
  Stat. Solidi B}\ }\textbf {\bibinfo {volume} {248}},\ \bibinfo {pages} {867}
  (\bibinfo {year} {2011})}\BibitemShut {NoStop}%
\bibitem [{\citenamefont {Bourgeois}\ \emph {et~al.}(2015)\citenamefont
  {Bourgeois}, \citenamefont {Jarmola}, \citenamefont {Siyushev}, \citenamefont
  {Gulka}, \citenamefont {Hruby}, \citenamefont {Jelezko}, \citenamefont
  {Budker},\ and\ \citenamefont {Nesladek}}]{Bourgeois2015}%
  \BibitemOpen
  \bibfield  {author} {\bibinfo {author} {\bibfnamefont {E.}~\bibnamefont
  {Bourgeois}}, \bibinfo {author} {\bibfnamefont {A.}~\bibnamefont {Jarmola}},
  \bibinfo {author} {\bibfnamefont {P.}~\bibnamefont {Siyushev}}, \bibinfo
  {author} {\bibfnamefont {M.}~\bibnamefont {Gulka}}, \bibinfo {author}
  {\bibfnamefont {J.}~\bibnamefont {Hruby}}, \bibinfo {author} {\bibfnamefont
  {F.}~\bibnamefont {Jelezko}}, \bibinfo {author} {\bibfnamefont
  {D.}~\bibnamefont {Budker}}, \ and\ \bibinfo {author} {\bibfnamefont
  {M.}~\bibnamefont {Nesladek}},\ }\href {\doibase 10.1038/ncomms9577}
  {\bibfield  {journal} {\bibinfo  {journal} {Nat Commun}\ }\textbf {\bibinfo
  {volume} {6}},\ \bibinfo {pages} {8577} (\bibinfo {year} {2015})}\BibitemShut
  {NoStop}%
\bibitem [{\citenamefont {Johnson}\ \emph {et~al.}(2015)\citenamefont
  {Johnson}, \citenamefont {Dolan}, \citenamefont {Grange}, \citenamefont
  {Trichet}, \citenamefont {Hornecker}, \citenamefont {Chen}, \citenamefont
  {Weng}, \citenamefont {Hughes}, \citenamefont {Watt}, \citenamefont
  {Auff{\`e}ves},\ and\ \citenamefont {Smith}}]{Johnson2015}%
  \BibitemOpen
  \bibfield  {author} {\bibinfo {author} {\bibfnamefont {S.}~\bibnamefont
  {Johnson}}, \bibinfo {author} {\bibfnamefont {P.~R.}\ \bibnamefont {Dolan}},
  \bibinfo {author} {\bibfnamefont {T.}~\bibnamefont {Grange}}, \bibinfo
  {author} {\bibfnamefont {A.~A.~P.}\ \bibnamefont {Trichet}}, \bibinfo
  {author} {\bibfnamefont {G.}~\bibnamefont {Hornecker}}, \bibinfo {author}
  {\bibfnamefont {Y.~C.}\ \bibnamefont {Chen}}, \bibinfo {author}
  {\bibfnamefont {L.}~\bibnamefont {Weng}}, \bibinfo {author} {\bibfnamefont
  {G.~M.}\ \bibnamefont {Hughes}}, \bibinfo {author} {\bibfnamefont {A.~A.~R.}\
  \bibnamefont {Watt}}, \bibinfo {author} {\bibfnamefont {A.}~\bibnamefont
  {Auff{\`e}ves}}, \ and\ \bibinfo {author} {\bibfnamefont {J.~M.}\
  \bibnamefont {Smith}},\ }\href
  {http://stacks.iop.org/1367-2630/17/i=12/a=122003} {\bibfield  {journal}
  {\bibinfo  {journal} {New J. Phys.}\ }\textbf {\bibinfo {volume} {17}},\
  \bibinfo {pages} {122003} (\bibinfo {year} {2015})}\BibitemShut {NoStop}%
\end{thebibliography}

\end{document}